\documentclass[epj]{svjour}

\usepackage{graphics}

\usepackage{latexsym}
\usepackage{axodraw}

\def\bq{\begin{eqnarray*}}
\def\eq{\end{eqnarray*}}
\def\l{\langle}
\def\r{\rangle}
\def\eps{\varepsilon}

\begin{document}
\title{Cancellation of infrared divergences at NNLO}
\author{Stefan Weinzierl
\thanks{Heisenberg fellow of the Deutsche Forschungsgemeinschaft}
}
\institute{Max-Planck-Institut f\"ur Physik (Werner-Heisenberg-Institut), F\"ohringer Ring 6, D-80805 M\"unchen, Germany}
\date{Received: date / Revised version: date}
%
\abstract{
Perturbative calculations at next-to-next-to-leading order
for multi-particle final states require a method to cancel
infrared singularities.
I discuss how to setup the subtraction method at NNLO.
} 
\maketitle
%

\section{Introduction}
\label{sect:intro}

The next generation of collider experiments will hunt
for the Higgs and other yet-to-be-discovered particles 
with increased luminosity and experimental precision.
The increased experimental precision has to be matched by an 
improvement in the accuracy of theoretical predictions.
Theoretical predictions are calculated as a power expansion in the coupling.
Higher precision is reached by including the next higher term 
in the perturbative expansion.
The experimental needs are numerical programs which yield predictions
for a wide range of observables.
Urgently needed are therefore fully differential
next-to-next-to-lead\-ing order (NNLO) programs.
Compared to certain specific NNLO prediction for inclusive observables,
these programs are flexible and allow
to take into account complicated detector geometries
and jet definitions.
The only requirement on the observable is infrared-safety.
At NNLO this implies that whenever a 
$n+1$ parton configuration $p_1$,...,$p_{n+1}$ becomes kinematically degenerate 
with a $n$ parton configuration $p_1'$,...,$p_{n}'$
we must have
\bq
{\cal O}_{n+1}(p_1,...,p_{n+1}) & \rightarrow & {\cal O}_n(p_1',...,p_n').
\eq
In addition, we must have in the double unresolved case (e.g. when
a $n+2$ parton configuration $p_1$,...,$p_{n+2}$ becomes kinematically degenerate 
with a $n$ parton configuration $p_1'$,...,$p_{n}'$)
\bq
{\cal O}_{n+2}(p_1,...,p_{n+2}) & \rightarrow & {\cal O}_n(p_1',...,p_n').
\eq
To construct such NNLO programs the following ingredients are needed:
\begin{description}
\item{-} The scattering amplitudes. This implies in particular for a NNLO
program the calculation of the relevant two-loop amplitudes.
There has been substantial progress in this field in the past years.
The state-of-the-art is that all two-loop-amplitudes, which are needed
most urgently, are now known
\cite{Bern:2000ie,Bern:2000dn,Anastasiou:2000kg,Anastasiou:2000ue,Anastasiou:2000mv,Anastasiou:2001sv,Glover:2001af,Bern:2001dg,Bern:2001df,Bern:2002tk,Garland:2001tf,Garland:2002ak,Moch:2002hm}.
\item{-} A NNLO program requires a method to cancel infrared divergences. 
Loop amplitudes, calculated in dimension\-al regularization, 
have explicit poles in the 
dimensional regularization parameter $\eps=2-D/2$,
arising from infrared singularities.
These poles cancel with similar poles arising from
amplitudes with additional partons, when integrated over phase space regions where
two (or more) partons become ``close'' to each other.
However, the cancellation occurs only after the integration over the unresolved phase space
has been performed and prevents thus a naive Monte Carlo approach for a fully exclusive
calculation.
It is therefore necessary to cancel first analytically all infrared divergences and to use
Monte Carlo methods only after this step has been performed.
\item{-} The final numerical computer program, which 
evaluates the remaining phase space integrals, requires
stable and efficient Monte Carlo methods
for this integration.
\end{description}
In this talk I focus on the cancellation of infrared divergences
\cite{Weinzierl:2003fx,Weinzierl:2003ra}.
In the next section I review general methods at NLO. In sect. \ref{sect:doubleunres} I discuss
the subtraction method at NNLO.
Sect. \ref{sect:oneloop} is devoted to one-loop amplitudes with one unresolved parton.

\section{A review of the subtraction method at NLO}
\label{sect:review}

\begin{figure}
\resizebox{0.5\textwidth}{!}{
  \includegraphics[115pt,650pt][410pt,720pt]{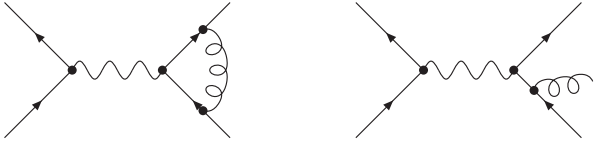}
}
\caption{Cancellation of divergences between virtual and real corrections at NLO.}
\label{fig:5}
\end{figure}

Infrared divergences occur already at next-to-leading order.
As a simple example two diagrams contributing to the NLO corrections
to $e^+ e^- \rightarrow \mbox{2 jets}$ 
are shown in fig. \ref{fig:5}.
The diagrams are divided into virtual and real corrections.
The virtual corrections contain the loop integrals and can have,
in addition to ultraviolet divergences, infrared divergences.
For one-loop amplitudes the IR divergences manifest themselves as
explicit poles in $\eps$ up to $1/\eps^2$.
For each IR divergence in the virtual corrections
there is a corresponding divergence with the opposite sign
in the real emission amplitude, obtained from the integration over 
the phase space region where some particles become
soft or collinear (e.g. unresolved).
In general, the Kinoshita-Lee-Nauenberg theorem
guarantees that any infrared-safe observable, when summed over all 
states degenerate according to some resolution criteria, will be finite.
However, the two contributions (virtual and real) live on different
phase spaces and prevent a naive Monte Carlo approach.
At NLO, general methods to circumvent this problem are known.
This is possible due to the universality of the singular behaviour
of the amplitudes in soft and collinear limits.
Examples are the phase-space slicing method
\cite{Giele:1992vf,Giele:1993dj,Keller:1998tf}
and the subtraction method
\cite{Frixione:1996ms,Catani:1997vz,Dittmaier:1999mb,Phaf:2001gc,Catani:2002hc}.
I briefly review the subtraction method here.
The NLO cross section is given as the sum of the virtual and real corrections:
\bq
\sigma^{NLO} & = & \int\limits_{n+1} d\sigma^R + \int\limits_n d\sigma^V.
\eq
If one can find an approximation term $d\sigma^A$ such that
\begin{itemize}
\item $d\sigma^A$ has the same point-wise singular behaviour in $D$ dimensions as $d\sigma^R$ itself, 
\item $d\sigma^A$ can be integrated analytically in $D$ dimensions over the 
one-parton subspace leading to soft and col\-linear divergences,
\end{itemize}
then one can add and subtract this term as follows:
\bq
\sigma^{NLO} 
& = & \int\limits_{n+1} \left( d\sigma^R - d\sigma^A \right) + \int\limits_n \left( d\sigma^V + \int\limits_1 
d\sigma^A \right).
\eq
Since by definition $d\sigma^A$ has the same singular behaviour as $d\sigma^R$, $d\sigma^A$
acts as a local counter-term and the combination $(d\sigma^R-d\sigma^A)$ is integrable
and can be evaluated numerically.
Secondly, the analytic integration of $d\sigma^A$ over the one-parton subspace will yield
the explicit poles in $\eps$ needed to cancel the corresponding poles in $d\sigma^V$.

\section{The subtraction method at NNLO}
\label{sect:doubleunres}

The following terms contribute at NNLO:
\bq
d\sigma_{n+2}^{(0)} & = & 
 \left( \left. {\cal A}_{n+2}^{(0)} \right.^\ast {\cal A}_{n+2}^{(0)} \right) 
d\phi_{n+2},  \nonumber \\
d\sigma_{n+1}^{(1)} & = & 
 \left( 
 \left. {\cal A}_{n+1}^{(0)} \right.^\ast {\cal A}_{n+1}^{(1)} 
 + \left. {\cal A}_{n+1}^{(1)} \right.^\ast {\cal A}_{n+1}^{(0)} \right)  
d\phi_{n+1}, \nonumber \\
d\sigma_n^{(2)} & = & 
 \left( 
 \left. {\cal A}_n^{(0)} \right.^\ast {\cal A}_n^{(2)} 
 + \left. {\cal A}_n^{(2)} \right.^\ast {\cal A}_n^{(0)}  
 + \left. {\cal A}_n^{(1)} \right.^\ast {\cal A}_n^{(1)} \right) d\phi_n,
\eq
where ${\cal A}_n^{(l)}$ denotes an amplitude with $n$ external partons and $l$ loops.
$d\phi_n$ is the phase space measure for $n$ partons.
Taken separately, each of these contributions is divergent.
Only the sum of all contributions is finite.
To render the individual contributions finite, one adds and subtracts suitable
pieces:
\bq
\lefteqn{
\l {\cal O} \r_n^{NNLO} = }
\nonumber \\
 & &
 \int \left( {\cal O}_{n+2} \; d\sigma_{n+2}^{(0)} 
             - {\cal O}_{n+1} \circ d\alpha^{(0,1)}_{n+1}
             - {\cal O}_{n} \circ d\alpha^{(0,2)}_{n} 
      \right) \nonumber \\
& &
 + \int \left( {\cal O}_{n+1} \; d\sigma_{n+1}^{(1)} 
               + {\cal O}_{n+1} \circ d\alpha^{(0,1)}_{n+1}
               - {\cal O}_{n} \circ d\alpha^{(1,1)}_{n}
        \right) \nonumber \\
& & 
 + \int \left( {\cal O}_{n} \; d\sigma_n^{(2)} 
               + {\cal O}_{n} \circ d\alpha^{(0,2)}_{n}
               + {\cal O}_{n} \circ d\alpha^{(1,1)}_{n}
        \right).
\eq
Here $d\alpha_{n+1}^{(0,1)}$ is a subtraction term for single unresolved configurations
of Born amplitudes.
This term is already known from NLO calculations.
The term $d\alpha_n^{(0,2)}$ is a subtraction term 
for double unresolved configurations.
Finally, $d\alpha_n^{(1,1)}$ is a subtraction term
for single unresolved configurations involving one-loop amplitudes.

To construct these terms the universal factorization properties of 
QCD amplitudes in unresolved limits are essential.
QCD amplitudes factorize if they are decomposed into primitive
amplitudes.
Primitive amplitudes are defined by
a fixed cyclic ordering of the QCD partons,
a definite routing of the external fermion lines through the diagram
and the particle content circulating in the loop.
\begin{figure}
\resizebox{0.5\textwidth}{!}{
  \includegraphics[100pt,600pt][545pt,705pt]{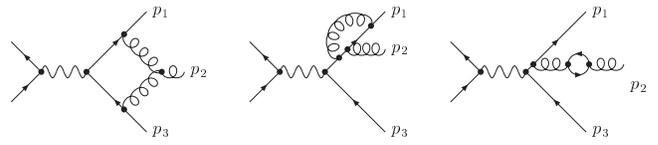}
}
\caption{Diagrams contributing to different primitive amplitudes.}
\label{fig:6}
\end{figure}
Fig. \ref{fig:6} shows three one-loop diagrams 
for $e^+ e^- \rightarrow \mbox{3 jets}$
contributing to different
primitive amplitudes.
One-loop amplitudes factorize in single unresolved limits as
\cite{Bern:1994zx,Bern:1998sc,Kosower:1999xi,Kosower:1999rx,Bern:1999ry,Catani:2000pi,Kosower:2003cz}
\begin{eqnarray}
\label{oneloopfactformula}
A^{(1)}_{n}
  & = &
  \mbox{Sing}^{(0,1)} 
  \cdot A^{(1)}_{n-1} +
  \mbox{Sing}^{(1,1)} \cdot A^{(0)}_{n-1}.
\end{eqnarray}
Tree amplitudes factorize in the double unresolved limits as
\cite{Berends:1989zn,Gehrmann-DeRidder:1998gf,Campbell:1998hg,Catani:1998nv,Catani:1999ss,DelDuca:1999ha,Kosower:2002su}
\bq
\label{factsing}
A^{(0)}_{n}
  & = &
  \mbox{Sing}^{(0,2)} \cdot A^{(0)}_{n-2}.
\eq
To discuss the term $d\alpha_n^{(0,2)}$ let us consider as an example
the Born leading-colour contributions to $e^+ e^- \rightarrow q g g \bar{q}$,
which contribute to the NNLO corrections to
$e^+ e^- \rightarrow \mbox{2 jets}$.
The subtraction term has to match all double and single unresolved 
configurations.
The double unresolved configurations are:\\
- Two pairs of separately collinear particles,\\
- Three particles collinear,\\
- Two particles collinear and a third soft particle,\\
- Two soft particles,\\
- Coplanar degeneracy.\\
The single unresolved configurations are:\\
- Two collinear particles,\\
- One soft particle.\\
It is convenient to construct $d\alpha_n^{(0,2)}$ as a sum over 
several pieces,
\bq
d \alpha^{(0,2)}_{n} & = & 
 \sum\limits_{\mbox{\tiny topologies $T$}} {\cal D}_{n}^{(0,2)}(T).
\eq
Each piece is labelled by a splitting topology.
\begin{figure}
\resizebox{!}{25mm}{
  \includegraphics[0pt,620pt][200pt,725pt]{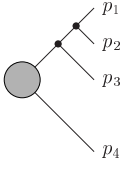}
}
\caption{Splitting topology.}
\label{fig:7}
\end{figure}
An example is shown in fig. \ref{fig:7}.
The term ${\cal D}_{n}^{(0,2)}(T)$ corresponding to the topology shown in
fig. \ref{fig:7} approximates singularities in $1/s_{12}$,
$1/(s_{12} s_{123})$ and part of the singularities in $1/s_{123}^2$.
Care has to be taken to disentangle correctly overlapping singularities
like $1/(s_{12}s_{23})$.
Details can be found in \cite{Weinzierl:2003fx}.

\section{One-loop amplitudes with one unresolved parton}
\label{sect:oneloop}

Apart from $d\alpha_n^{(0,2)}$ also the term
$d\alpha_n^{(1,1)}$, which approximates one-loop amplitudes 
with one unresolved parton, is needed at NNLO.
If we recall the factorization formula (\ref{oneloopfactformula}),
this requires as a new feature 
the approximation of the one-loop singular function
$\mbox{Sing}^{(1,1)}$.
The corresponding subtraction term is proportional to the 
one-loop $1\rightarrow 2$ splitting function 
${\cal P}^{(1,1)}_{(1,0)\; a \rightarrow b c}$.
An example is the leading-colour part for the
splitting $q \rightarrow q g$:
\bq
\lefteqn{
{\cal P}^{(1,1)}_{(1,0)\; q \rightarrow q g, lc, corr} 
 =  
     - \frac{11}{6\eps} {\cal P}^{(0,1)}_{q \rightarrow q g},
 +
 S_\eps^{-1} c_\Gamma \left( \frac{-s_{ijk}}{\mu^2} \right)^{-\eps} 
  y^{-\eps}
 }
 \nonumber \\
 & &
     \left\{ 
         g_{1, corr}(y,z) \; {\cal P}^{(0,1)}_{q \rightarrow q g}
         + f_2 \frac{2}{s_{ijk}} \frac{1}{y} p\!\!\!/_{e} 
               \left[ 1 - \rho \eps (1-y) (1-z) \right]
     \right\}.
\eq
This term depends on the correlations among the remaining hard partons.
If only two hard partons are correlated, $g_{1}$ is given by
\bq
\lefteqn{
g_{1, intr}(y,z) 
 = 
  - \frac{1}{\eps^2} 
 \left[ \Gamma(1+\eps) \Gamma(1-\eps) \left( \frac{z}{1-z} \right)^\eps 
        + 1 
\right. } \nonumber \\ & & \left.
        - (1-y)^\eps z^\eps \; {}_2F_1\left( \eps, \eps, 1+\eps; (1-y)(1-z) \right) \right].
\hspace{10mm}
\eq
For the integration of the subtraction terms 
over the unresolved phase space all occuring integrals are reduced to
standard integrals of the form
\bq
\lefteqn{
\int\limits_0^1 dy \; y^a (1-y)^{1+c+d} \int\limits_0^1 dz \; z^c (1-z)^d \left[ 1 -z(1-y)\right]^e
} \nonumber \\
\lefteqn{
  {}_2F_1\left( \eps, \eps; 1+\eps; (1-y) z \right)
 = }
 \nonumber \\ & &
 \frac{\Gamma(1+a) \Gamma(1+d) \Gamma(2+a+d+e) \Gamma(1+\eps)}{\Gamma(2+a+d) \Gamma(\eps) \Gamma(\eps)}
 \nonumber \\ & &
 \sum\limits_{j=0}^\infty 
 \frac{\Gamma(j+\eps) \Gamma(j+\eps) \Gamma(j+1+c)}
      {\Gamma(j+1) \Gamma(j+1+\eps) \Gamma(j+3+a+c+d+e)}.
\eq
The result is proportional to a hyper-geometric functions 
${}_4F_3$ with unit argument and can be
expanded into a Laurent series in $\eps$ 
with the techniques of \cite{Moch:2001zr,Weinzierl:2002hv}.
For the example discussed above one finds after integration
\cite{Weinzierl:2003ra}:
\bq
\lefteqn{
{\cal V}^{(1,1)}_{(1,0)\; q \rightarrow q g, lc, intr} = 
 - \frac{1}{4\eps^4}
 - \frac{31}{12 \eps^3}
 + \left( -\frac{51}{8} - \frac{1}{4} \rho + \frac{5}{12} \pi^2 
\right. }
 \nonumber \\
 & & \left.
          - \frac{11}{6} L
   \right) \frac{1}{\eps^2} 
 + \left( - \frac{151}{6} - \frac{55}{24} \rho 
          + \frac{145}{72} \pi^2 + \frac{15}{2} \zeta_3
          - \frac{11}{4} L 
 \right. \nonumber \\ & & \left.
          - \frac{11}{12} L^2 
   \right) \frac{1}{\eps}
 - \frac{1663}{16} - \frac{233}{24} \rho 
 + \frac{107}{16} \pi^2 + \frac{5}{12} \rho \pi^2 
 + \frac{356}{9} \zeta_3 
 \nonumber \\ & & 
 - \frac{1}{72} \pi^4
 - \frac{187}{24} L - \frac{11}{12} \rho L + \frac{55}{72} \pi^2 L
 - \frac{11}{8} L^2 - \frac{11}{36} L^3
 \nonumber \\
 & &
 + i \pi \left[
            - \frac{1}{4 \eps^3}
            - \frac{3}{4 \eps^2}
            + \left( - \frac{29}{8}
                     - \frac{1}{4} \rho + \frac{\pi^2}{3} \right) \frac{1}{\eps}
            - \frac{139}{8} - \frac{11}{8} \rho 
 \right. \nonumber \\ & & \left.
            + \pi^2 + \frac{15}{2} \zeta_3 
     \right]
 + {\cal O}(\eps),
\eq
where $L = \ln(s_{ijk}/\mu^2)$.
The parameter $\rho$ specifies the variant of dimensional regularization:
$\rho  = 1$ in the conventional or 't Hooft-Veltman scheme and 
$\rho=0$ in a four-dimensional scheme.

\section{Outlook}
\label{sect:summary}

In this talk I reported on the subtraction method to cancel infrared 
divergences at NNLO.
The set-up involves two new types of subtraction terms, $d\alpha^{(0,2)}_{n}$
and $d\alpha^{(1,1)}_{n}$.
The former approximates double unresolved configurations of tree amplitudes
with $n+2$ partons, whereas the latter approximates one-loop amplitudes
in single unresolved limits.
Decomposing the QCD amplitudes into partial and primitive amplitudes,
the appropriate subtraction terms have been constructed.
Furthermore, the analytic integration over the unresolved phase space has been
performed for all terms contributing to $d\alpha^{(1,1)}_{n}$.
Once the corresponding analytic integration has been done for $d\alpha^{(0,2)}_{n}$
the subtraction method at NNLO is complete and can used for fully 
differential programs at NNLO.

%
%

\end{document}